\documentclass[aps,showpacs,twocolumn]{revtex4}

\usepackage{amssymb}
\usepackage{amsmath}
\usepackage{epsf}
\usepackage{graphicx}
\usepackage{psfig}
\usepackage[dvips]{epsfig}
\usepackage{dcolumn}
\usepackage{bm}

\begin{document}

\title{Photon statistics dispersion in excitonic composites}

\author{G. Ya. Slepyan}
\author{S. A. Maksimenko}
\email{maksim@bsu.by}
\affiliation{Institute for Nuclear Problems, Belarus State
University, 11 Bobruiskaya Str., 220050 Minsk, BELARUS}


\begin{abstract}
Linear media are predicted to exist whose relative permiability is
an operator in the space of quantum states of light. Such media
are characterized by a photon statistics--dependent  refractive
index. This indicates a new type of optical dispersion -- the
photon statistics dispersion.  Interaction of quantum light with
such media modifies the photon number distribution and, in
particular, the degree of coherence of light. An excitonic composite --
a collection of noninteracting quantum dots -- is considered as a
realization of the medium with the photon statistics dispersion. Expressions are derived for generalized plane waves in an excitonic composite and input--output relations for a planar layer of the material. Transformation rules for different photon initial states are analyzed. Utilization of the photon statistics dispersion in potential quantum--optical devices is discussed.
\end{abstract}

\pacs{42.50.Nn, 42.50.Ct, 78.67.Hc, 42.50.Pq}

\maketitle

\section{Introduction}

The search
and investigation of novel physical processes transforming the
photon number distribution is one of the most important problems of
quantum optics \cite{Scully_b01}. Processes of that type known
so far can be separated into two groups. The first group
comprises different interference effects, such as two--beam
interference of photons in a beam--splitter \cite{Scully_b01}. In some schemes, the use of ancillary photons and conditional detection enables the realization of effective nonlinear interaction by means of linear optics \cite{Knill_Nature_01}. Besides their fundamental importance, such processes have tremendous potential for  a quantum computing. In particular, such transformation of photon statistics as nonlinear sign shift 
\cite{Knill_Nature_01,Sanaka_PRL_04}  and arbitrary photon number state filtering \cite{Sanaka_PRL_06} have been realized.
The manifestation of phase--dependent
photon statistics in the mixed field of a narrow band two-photon
source and coherent field has been reported in Ref. \cite{Lu_02}. Photon antibunching as well as photon bunching have been observed. One can expect that this scheme should be able to produce effectively sub-Poissonian photon statistics.

The second group is formed by the
processes in systems with nonlinearity and gain, where the
transformation occurs due to the strong light--matter coupling \cite{Scully_b01}. A major development is that of single--photon sources \cite{Lounis_05}. The strong light--matter coupling enables different methods of generation of the sub-Poissonian photon statistics, such as micromaser \cite{Scully_b01}, single--atom resonance fluorescence \cite{Short_PRL_83}, single--trapped--atom laser \cite{Keever_Nature_03}, and cavity--QED laser \cite{Choi_PRL_06}. In Ref. \cite{Law_PRL_96}, a method has been proposed of the transformation of a vacuum state of cavity mode into arbitrary quantum state of light by means of succession of strong atom--field interaction processes inside the cavity.

In the present paper, we present an
opportunity of a new type:  transformation of the photon statistics  in a
\textit{linear} homogeneous medium those refractive index depends on the photon number distribution.  As a physical realization of a medium with the photon
statistics--dependent refractive index, we propose a collection of noninteracting quantum dots
(QDs) embedded in a host semiconductor. This composite is called an \emph{excitonic
composite} \cite{Slepyan_NATO_03,Maksimenko_ENN}.

The optical and electronic properties of QDs are currently in vogue, due to promise in 
semiconductor device physics \cite{bimberg_b99} and, in
particular, for the quantum information storage and processing
\cite{Yamamoto_b,Michler_b03}. The application of  QDs as
potential quantum--light emitters
\cite{Lounis_05,Michler_b03,Michler_00,Santori_PRL_01,Moreau_PRL_02,Regelman_01} is now
intensively discussed. In particular, Ref. \cite{Regelman_01}
demonstrates a quantum light source of multicolor photons with
tunable statistics. The coupling of two QDs has been demonstrated as an effective means of emitted photons antibunching \cite{Gerardot_PRL_05}. 

The peculiar property of a QD exposed to electromagnetic field is the pronounced role of the dipole--dipole electron--hole interaction. Phenomenologically,
this interaction can be introduced trough the \emph{local fields} \cite{Slepyan_pra02,Maksimenko_HN04,Ajiki_02,Goupalov_03}. Obviously, in quantum
electrodynamics the dipole--dipole interaction is due to the exchange between electrons and holes by virtual vacuum photons \cite{Lewenstein_94}. Refs. \cite{Slepyan_pra02,Maksimenko_HN04}
predict a fine structure of the absorption (emission) line in a QD
interacting with quantum light. Instead of a single line at
the exciton transition frequency $\omega_0$, a doublet appears
with one component blue (red) shifted  by a value
$\Delta\omega$. The fine structure has no analog in classical
electrodynamics. The value of the shift depends only on the QD
shape, while the intensities of components are completely
determined by the light statistics. In the limiting cases of
classical light and single Fock states, the doublet reduces to
a singlet shifted in the former case and unshifted in  the latter
one. In heterogeneous mediums comprising  regular or irregular
arrays of QDs embedded in a semiconducting host, this
mechanism is responsible for the photon statistics dispersion mentioned earlier.

The electromagnetic response properties of a composite medium comprising electrically small (of dimension much smaller than the
wavelength) inclusions
exposed to  classical light can be modeled within the
effective--medium approach modified
\cite{Slepyan_NATO_03,Maksimenko_ENN} to include specific
properties of excitons as resonant states with discrete energy
spectrum. To solve the problem of the quantum light interaction
with a composite medium, an adequate quantization technique
must be developed. Conventionally, particular models of the
classical light interaction with homogeneous media are exploited
for this purpose. Nonhomogeneites are implemented into models by means of homogenization
procedure on the classical stage. However, the subsequent  replacement of classical fields by corresponding
field operators often leads to a fundamental problem, which is the lack of
correct commutation relations for the field operators. To overcome
the problem, some modifications of the quantization scheme
have been proposed, such as, for instance, the noise
current concept \cite{Welsch_b01}. Nevertheless, in all cases to
our knowledge, the constitutive parameters in classical and
quantum optics are assumed to be identical. 

Here we propose an
alternative approach: first we consider the interaction of quantum
light with an isolated scatterer (QD) and then we develop a
corresponding homogenization technique. The approach exploits the well-known technique \cite{Goldberger,Landau_QM} linking the macroscopic refractive index to the
forward scattering amplitude $f(0)$ by
\begin{equation}
\label{index}
    n^2=1+\frac{4\pi \rho}{k^2}
f(0)\,,
\end{equation}
where $k=\omega/c$ is the vacuum wavenumber, $c$ is vacuum speed of light, and $\rho$ is the density
of scatterers in the medium. Within the approach, the quantum field operators satisfy the correct commutation relations automatically. Note that the approach is applicable
not only to photons but to quantum fields of other origination
(atoms and atomic nuclei, neutrons, etc.) and is in the basis of
nuclear optics of polarized media \cite{Baryshevskii}.

The paper is arranged
as follows. In Sec. II we present a model of the single QD--quantum light interaction and evaluate the forward scattering amplitude operator. The fundamentals of quantum optics of excitonic composites --- constitutive relations for field operators, planewave  solution of Maxwell equations, input--output relations for a planar layer --- are formulated in Sec. III. Sec. IV is  devoted to the photon statistics dispersion. A general formulation of the transformation law for the light density operator and the second--order coherence correlation function are set up and some particular examples of different initial quantum states are considered. A summary of the work and outlook are given in Sec. V.  The Appendix provides some details of derivations presented in Sec. II.

\section{Interaction of quantum light with a single QD}

\subsection{General relations}

We start with the evaluation of the forward scattering amplitude $f(0)$ of quantum light interacting with a single QD.  As has been pointed out by Cho \cite{Cho_b03}, in the formulation of Hamiltonian of the interacting system ''electromagnetic field + condensed matter'', the decomposition of the system into
matter and field is ambiguous. In the approach called scheme [A] \cite{Cho_b03}, only the transverse part of  the total field
--- radiation field --- is attributed to electromagnetic field.
Charge carriers are assumed to be interacting via longitudinal near field which thereby is attributed to the continuum. In such a scheme, the free
electromagnetic field is described in terms of the photon creation/annihilation operators $\hat{a}^\dag,~\hat{a}$ while the QD continuum is described by the
quantum field operators $\widehat\Psi(\mathbf{r})$ and $\widehat\Psi^\dag(\mathbf{r})$. These quantum fields are fermionic \cite{Cho_b03} and, consequently,
satisfy the anticommutation relations:
    \begin{equation}
    \begin{array}{lcl}
    [\widehat{\Psi}^\dag(\mathbf{r}),\widehat
    {\Psi}(\mathbf{r}')]_{+}
    &=&
    \delta (\mathbf{r} -
    \mathbf{r}')\,,
    \\ \rule{0in}{5ex}
    [\widehat{\Psi}(\mathbf{r}),\widehat {\Psi}(\mathbf{r}')]_{+}
    &=& [\widehat {\Psi}^\dag(\mathbf{r}),\widehat
    {\Psi}^\dag({\mathbf{r}}')]_{+} = 0\,.
    \end{array}
    \label{13:commut}
    \end{equation}
Accordingly \cite{Cho_b03},  the Hamiltonian of a QD exposed to quantum electromagnetic field can be written as
\begin{equation}
\label{total_H}
    {\cal H}={\cal H}_0+{\cal H}_\mathrm{ph}^{}
+{\cal H}_\mathrm{I}^{}+{\cal H}_\mathrm{C}^{}\,.
\end{equation}
Here ${\cal H}_{0}$ is the Hamiltonian of the charge carriers free motion.  The term ''free'' implies the carriers to be spatially confined in the QD but free of
the influence of electromagnetic field and interparticle interaction.  The Hamiltonian 
\begin{equation}
\label{H_ph}
    {\cal H}_\mathrm{ph}^{}
    =\int\limits_0^\infty\hbar\omega\,
    \hat{a}_\omega^\dag\hat{a}_\omega\, d\omega\,
\end{equation}
describes  photons in free space in the single--mode approximation \cite{Scully_b01}. The index $\omega$ ascribes  the creation/annihilation operators to photon of a frequency $\omega$.  The
creation/annihilation operators satisfy the commutation relation $[\hat{a}_\omega,\,\hat{a}_{\omega'}^\dag]=\delta(\omega-\omega')$. The light--matter interaction Hamiltonian ${\cal
H}_{\mathrm{I}}^{}$ in the length gauge \cite{Scully_b01} is given  by
\begin{eqnarray} \label{H_I}
    {\cal H}_\mathrm{I}^{}&=& - \frac{1}{2}\int\limits_{V}(\widehat{\bm{\mathcal{P}}}\widehat{\bm{\mathcal{E}}}_\mathrm{0}+
    \widehat{\bm{\mathcal{E}}}_\mathrm{0}\widehat{\bm{\mathcal{P}}})\, d^3\mathbf{r}\,,
         \\
         \noalign{\hbox{where}}\rule{0in}{3ex}
    \widehat{\bm{\mathcal{P}}}(\mathbf{r})&=& e\,\widehat\Psi^\dag(\mathbf{r})\,\mathbf{r}\,
    \widehat {\Psi}(\mathbf{r})
\end{eqnarray}
is the time--domain polarization operator, $e$ is the electron charge,  $V$ is the QD volume and $\widehat{\bm{\mathcal{E}}}_0$ is the free electric field
operator. Besides the Heisenberg representation  of operators $\widehat{\bm{\mathcal{E}}}_0(\mathbf{r})$ and $\widehat{\bm{\mathcal{P}}}(\mathbf{r})$,
further we shell use their Fourier transforms with respect to time. In particular, the electric field operator can be represented by the superposition $\widehat{\bm{\mathcal{E}}}_0
(\mathbf{r})= \widehat{\bm{\mathcal{E}}}{}^{(+)}_0 (\mathbf{r})+ \widehat{\bm{\mathcal{E}}}{}^{(-)}_0 (\mathbf{r})$, where superscripts $(\pm)$ mark
contributions from positive and negative frequencies, respectively. Thus, the time--domain operator $\widehat{\bm{\mathcal{E}}}{}^{(+)}_0 (\mathbf{r})$ is
associated with the frequency--domain field by the positive semi--axis Fourier transform:
\begin{equation}
\label{E_+}
\widehat{\bm{\mathcal{E}}}{}^{(+)}_0 (\mathbf{r})=
    \int_o^\infty
    \widehat{\mathbf{E}}_0(\mathbf{r},\omega)d\omega\,.
\end{equation}
The operators $\widehat{\bm{\mathcal{E}}}{}^{(-)}_0 (\mathbf{r})$ and $\widehat{\bm{\mathcal{E}}}{}^{(+)}_0 (\mathbf{r})$ are hermitian conjugates.
In the single--mode approximation
\begin{equation}
\label{E_0}
    \widehat{\bf E}_0(\mathbf{r},\omega)=i\sqrt{\frac{\hbar k}{A}}\,
    {\bf e}_y\, \hat{a}_\omega\,e^{ikz}\,,
\end{equation}
where  ${\bf e}_y$ is the unit polarization vector of free photons and $A$ is the normalization cross--section in the $xy$ plane. The Fourier transforms of the
polarizability $\widehat{\bm{\mathcal{P}}}(\mathbf{r})$ and the electric displacement $\widehat{\bm{\mathcal{D}}}(\mathbf{r})$ are introduced in the same
manner.

The Hamiltonian  ${\cal H}_\mathrm{C}^{}$ in (\ref{total_H}) describes the interaction of particles in the QD. The retardation in that interaction can be ignored in
view of the electrical smallness of the QD. Then,  in the Hartree--Fock--Bogoliubov approximation, the Hamiltonian ${\cal H}_\mathrm{C}^{}$ is represented by ${\cal
H}_\mathrm{C}^{}={\cal H}_\mathrm{C1}^{}+{\cal H}_\mathrm{C2}^{}$, where
\begin{eqnarray}
\label{HC_1}
{\cal H}_\mathrm{C1}^{}= e^2\int\limits_V\!\!\int\limits_V
    \widehat\Psi^\dag(\mathbf{r})\,\widehat{\Psi}(\mathbf{r})\,
    \langle \widehat\Psi^\dag(\mathbf{r}')
    \widehat {\Psi}(\mathbf{r}')\rangle
    \frac{d^3\mathbf{r}\,d^3\mathbf{r}'}{|\mathbf{r}-\mathbf{r}'|}\,,
    \\\rule{0in}{4ex}
\label{H_C2}
    {\cal H}_\mathrm{C2}^{}= -e^2\int\limits_V\!\!\int\limits_V
    \widehat\Psi^\dag(\mathbf{r}')\,\widehat{\Psi}(\mathbf{r})\,
    \langle \widehat\Psi^\dag(\mathbf{r})
    \widehat {\Psi}(\mathbf{r}')\rangle
    \frac{d^3\mathbf{r}\,d^3\mathbf{r}'}{|\mathbf{r}-\mathbf{r}'|}\,.
\end{eqnarray}
The term ${\cal H}_\mathrm{C1}^{}$  accounts for the direct interaction of particles while the term ${\cal H}_\mathrm{C2}^{}$ defines the exchange
interaction. Here and hereafter the angular brackets $\langle\dots\rangle$ stand for the operator's mean value. Note that terms (\ref{HC_1}) and (\ref{H_C2})
comprise both static Coulomb interactions and local fields (dynamic Coulomb forces). As the next step, we reformulate the Hamiltonian (\ref{total_H}) in order
to isolate the contribution of the local  field.

The quantum field operators $\widehat\Psi(\mathbf{r})$ and $\widehat\Psi^\dag(\mathbf{r})$
are related to real particles in QDs --- atoms in crystalline lattice. The introduction of quasi--particles supposes averaging over the unit cell and transition to
envelopes. In accordance with the envelope technique \cite{Hang_b94}, the field operators can be represented by
\begin{equation}
    \widehat{\Psi}(\mathbf{r})=\sqrt{{V}/{N}}\,\sum_{n,q}\hat{\xi}_n(\mathbf{R}_q)
    w_{nq}(\mathbf{r})\,,
    \label{3:quantum_fields}
    \end{equation}
where  $N$ is a total number of atoms in the QD, $w_{nq}(\mathbf{r})$ is the Wannier function in the $q$-th point of the QD crystalline lattice, and
$\mathbf{R}_q$ is the radius--vector of that point, the index $n$ stands for the collection of quantum numbers characterizing electronic states in the QD,
and $\hat{\xi}_n(\mathbf{R}_q)$ is the quantum field envelope satisfying fermionic anticommutative relations (\ref{A:anticommut}). 

Next we substitute
(\ref{3:quantum_fields}) into (\ref{HC_1}) and (\ref{H_C2}) and carry out the averaging over the unit cell (see Appendix for details). Transition to the
continuous field of envelopes is realized by means of the replacement (\ref{A:eq12}); as a result, the Hamiltonian ${\cal H}_\mathrm{C}^{}$ is expressed by the
superposition ${\cal H}_\mathrm{C}^{}={\cal H}_\mathrm{0C}^{}+\Delta{\cal H}$, where
\begin{widetext}
\begin{eqnarray}
\label{HC_0}
{\cal H}_\mathrm{0C}^{}&=& e^2\sum_{m,n}\int\limits_V\!\!\int\limits_V
    \Bigl[\hat\xi_n(\mathbf{r})\,\hat{\xi}_n(\mathbf{r})\,
    \langle \hat\xi_m(\mathbf{r}')\hat {\xi}_m(\mathbf{r}')\rangle
    -\hat\xi_m^\dag(\mathbf{r})\,\hat{\xi}_n(\mathbf{r}')\,
    \langle \hat\xi_n^\dag(\mathbf{r})\hat{\xi}_m(\mathbf{r}')\rangle\Bigr]
    \frac{d^3\mathbf{r}\,d^3\mathbf{r}'}{|\mathbf{r}-\mathbf{r}'|}\,,
    \\\rule{0in}{4ex}
\label{Delta_Ha}
    \Delta{\cal H}&=& \frac{1}{2}\sum_{\substack{m,m' \\ n,n'}}\int\limits_V\!\!\int\limits_V
    \bm{\mu}_{nn'}\underline{G}(\mathbf{r}-\mathbf{r}')\bm{\mu}_{mm'}
    \Bigl[\hat\xi_n^\dag(\mathbf{r})\,\hat{\xi}_{n'}(\mathbf{r})\,
    \langle \hat\xi_m^\dag(\mathbf{r}')\hat {\xi}_{m'}(\mathbf{r}')\rangle
    -\hat\xi_m^\dag(\mathbf{r}')\,\hat{\xi}_{n'}(\mathbf{r})\,
    \langle \hat\xi_n^\dag(\mathbf{r})\hat{\xi}_{m'}(\mathbf{r}')\rangle\Bigr]
    \,{d^3\mathbf{r}\,d^3\mathbf{r}'}\,.
\end{eqnarray}
\end{widetext}
Here $\underline{G}(\mathbf{r})$ is the electrostatic  Green's tensor of free space and $\bm{\mu}_{nn'}$ is the transition dipole moment (see Eq. (\ref{A:eq9})
and  following comment). The term (\ref{HC_0})  describes the influence of static Coulomb forces while the Hamiltonian $\Delta{\cal H}$ (\ref{Delta_Ha})
accounts for the local fields.

\subsection{Model Hamiltonian}
\label{Hamiltonians}

For further analytical progress, we use a set of simplifications. First, we restrict the consideration to two level model of  exciton in QD,  with $|g\rangle $ and
$|e\rangle $ as  its ground  and excited states, and neglect intraband transitions, i.e., we let $\bm{\mu}_{nn}=0$. Exciton is assumed to be strongly
confined in the QD; under that assumption the wavefunctions of states $|g\rangle $ and $|e\rangle $ are the same \cite{Hang_b94}. Denoting them as
$\xi(\mathbf{r})$ we present the quantum field envelopes by $\hat\xi_\mathrm{g,e}=\hat{a}_\mathrm{g,e}\xi(\mathbf{r})$ and
$\hat\xi_\mathrm{g,e}^\dag=\hat{a}_\mathrm{g,e}^\dag\xi^*(\mathbf{r})$, where $\hat{a}_\mathrm{g,e}^\dag$ and $\hat{a}_\mathrm{g,e}$ are the operators of the
electron creation and annihilation in the ground and excited states, respectively.   Then we omit the term ${\cal H}_\mathrm{0C}^{}$ which accounts for the
Coulomb interactions and is small in the strong confinement regime \cite{Hang_b94}.   Lastly, in order to eliminate fast oscillations, we make use of the
rotating--wave approximation \cite{Scully_b01}. As a result, we come to the light--matter Hamiltonian (\ref{total_H}) with ${\cal
H}_{\mathrm{C}}^{}=\Delta{\cal H}$ and
\begin{eqnarray}
    \label{H_0A}
    {\cal H}_{0}^{} &=&\frac{1}{2}\hbar\omega_0\hat\sigma_z^{}\,\\ \rule{0in}{3ex}
        \label{H_IA}
        {\cal H}_\mathrm{I}^{}&=&\hbar\int\limits_0^\infty\Bigl(g_\omega\hat\sigma_+^{}\hat{a}_\omega+
        g_\omega^*\hat\sigma_-^{}\hat{a}_\omega^\dag\Bigr)\, d\omega\,,\\ \rule{0in}{3ex}
    \label{Delta_H}
    \Delta{\cal H}&=&\hbar\Delta\omega(
    \hat\sigma_-^{}\langle\hat\sigma_+^{}\rangle+\hat\sigma_+^{}\langle\hat\sigma_-^{}\rangle)\,,
\end{eqnarray}
where $\omega_0=(\varepsilon_\mathrm{e}-\varepsilon_\mathrm{g})/\hbar$ is  the resonant frequency of the transition in QD and  $\varepsilon_\mathrm{g,e}$ are
the energy eigenvalues in the ground and excited states, respectively, $\hat\sigma_{z,\pm}^{}$ are the Pauli pseudospin operators. The quantity
\begin{equation}
g_\omega=-i(\bm{\mu}\cdot{\bf e}_{y})   \sqrt{\frac{\omega}{\hbar c A}}
\end{equation}
 is the coupling factor for photons and carriers in QD; $\bm{\mu}=(\bm{\mu}_{\mathrm{eg}}\cdot{\bf e}_{y})\,{\bf e}_{y}=
(\bm{\mu}_{\mathrm{ge}}^*\cdot{\bf e}_{y})\,{\bf e}_{y}$. The depolarization shift
\begin{equation}
    \label{delta_w}
    \Delta\omega=\frac{4\pi}{\hbar V}(\bm{\mu}\cdot\tilde{\underline{N}}    \bm{\mu})
\end{equation}
characterizes the role of local fields inside QD. For a QD exposed to a classical electromagnetic field, the depolarization shift $\Delta\omega$ has been introduced in Refs. \cite{Slepyan_99a,Maksim_00a}. The depolarization tensor is given by \cite{Maksimenko_HN04}
 \begin{equation}
\label{1:depol_tens0} \tilde{\underline{N}}=-\frac{V}{4\pi }
    \int\limits_{V}\!\!\int\limits_{V}|\xi(\mathbf{r})|^2\,|{\xi}(\mathbf{r}')|^2
    \nabla_\mathbf{r}\otimes\nabla_\mathbf{r}
    \left(\frac{1}{|\mathbf{r}-\mathbf{r}'|}\right)
    d^3\mathbf{r}\,d^3\mathbf{r}'\,,
 \end{equation}
where $\nabla_\mathbf{r}\otimes\nabla_\mathbf{r}$ is  the operator dyadic acting on variables $\mathbf{r}$. Physical meaning of the local--field correction
becomes apparent if we rewrite the total interaction Hamiltonian ${\cal H}_\mathrm{I}^{}+{\cal H}_\mathrm{C}^{}\equiv{\cal H}_\mathrm{I}^{}+ \Delta{\cal H}$ in
the form as follows:
\begin{eqnarray} \label{H_I+D}
    {\cal H}_\mathrm{I}^{}+\Delta{\cal H}=-\frac{1}{2}\int\limits_{V}(\widehat{\bm{\mathcal{P}}}\widehat{\bm{\mathcal{E}}}_\mathrm{L}+
    \widehat{\bm{\mathcal{E}}}_\mathrm{L}\widehat{\bm{\mathcal{P}}})\, d^3\mathbf{r}\,,
\end{eqnarray}
where the local field operator $\widehat{\bm{\mathcal{E}}}_\mathrm{L}$ is related to the free electric field operator by
\begin{eqnarray}
\label{screen}
    \widehat{\bm{\mathcal{E}}}_\mathrm{L}(\mathbf{r})= \widehat{\bm{\mathcal{E}}}_\mathrm{0}-
    \frac{4\pi}{V}\Bigl(\tilde{\underline{N}}\bm{\mu}\langle\hat\sigma_+^{}\rangle+ \mathrm{H.c.}\Bigr) \,.
\end{eqnarray}
This relation demonstrates that the local field manifests itself  as the QD depolarization, i.e., the screening  of external field by charges induced on the QD
surface. The depolarization field defined by the second term in the right side of Eq. (\ref{screen}) is a classical field and is expressed by a $C$--number. The
result follows from that  the depolarization field  in the quasi--static approximation is longitudinal and thus is not quantized. That is, this field is not
expressed in terms of photonic operators and, consequently, does not experience quantum fluctuations. A similar situation has been considered in Ref.
\cite{Fleichhauer_pra_99} under the analysis of  Lorenz--Lorentz correction for quantum light in optically dense media.

Any  QD is essentially a multilevel system. However, the joint
contribution of all transitions lying far away from a given
resonance can be approximated by a nonresonant relative
permittivity $\varepsilon_h$.  The host semiconductor's relative
permittivity is also assumed to be equal to $\varepsilon_h$. For
analytical tractability, let $\varepsilon_h$ be
frequency--independent and real--valued. This allows us to put
$\varepsilon_h=1$ without loss of generality. The substitutions
$
  c\to c/\sqrt{\varepsilon_h}, ~\,{\bm \mu}\to {\bm \mu}/\sqrt{\varepsilon_h}
$
in the final expressions will restore the case $\epsilon_h\neq 1$.

\subsection{Equations of motion}

In the previous sections, we have formulated the light--matter interaction Hamiltonian (\ref{total_H}) and defined all its components in terms of the operators $\hat{a}_\omega$, $\hat{a}_\omega^\dag$, $\hat\sigma_\pm^{}$ and $\hat{\sigma}_z$ by Eqs. (\ref{H_ph}),
(\ref{H_0A})--(\ref{Delta_H}). In the Heisenberg representation, the equations of motion for operators $\hat{a}_\omega$, $\hat\sigma_-^{}$ and $\hat{\sigma}_z$
are given by
\begin{eqnarray}
        \label{motion_1}
        &&\frac{\partial}{\partial t}\hat\sigma_-^{}=-i\omega_0\hat\sigma_-^{}+
        \hat{\sigma}_z\int\limits_0^\infty g_\omega  \hat{a}_\omega \,d\omega+
        i\Delta\omega\hat{\sigma}_z\langle\hat\sigma_-^{}\rangle\,,\qquad
        \\ \rule{0in}{3ex}
        \label{motion_2}
        &&\frac{\partial}{\partial t}\hat{a}_\omega = - i\omega\hat{a}_\omega-ig_\omega\hat\sigma_-^{}\,,
        \\ \rule{0in}{3ex}
        \label{motion_3}
        &&\frac{\partial}{\partial t}\hat\sigma_z=
        2\int\limits_0^\infty g_\omega
        (\hat{a}_\omega^\dag \hat\sigma_-^{}-\hat\sigma_+^{}\hat{a}_\omega)\,d\omega
        \cr \rule{0in}{3ex}
        &&\qquad\qquad\qquad+2\hbar\Delta\omega(\hat\sigma_-^{}\langle\hat\sigma_+^{}\rangle
    +\hat\sigma_+^{}\langle\hat\sigma_-^{}\rangle)\,,
\end{eqnarray}
Let at the initial time $t=-\infty$ the QD be in the ground state $|g\rangle $ ($\langle\hat\sigma_z(-\infty)\rangle=-1$) and be exposed to an arbitrary
quantum state of light. The interaction is assumed to be weak, i.e., only first--order terms in $g_\omega$ are of importance. From Eq. (\ref{motion_1}) follows
that $\hat\sigma_{\pm}^{}\,$ and $\langle\hat\sigma_{\pm}^{}\rangle\sim O(g_\omega)$, which means that ${\partial\hat\sigma_z}/{\partial t}\sim
O(|g_\omega|^2)$ in (\ref{motion_3}). Consequently, the weak interaction model allows us to put $\hat\sigma_z(t)=\hat\sigma_z(-\infty)$ in the further
analysis. Since the final expressions will obligatorily involve averaging over the ground state, i.e., the quantity
$\langle\hat\sigma_z(t)\rangle\approx\langle\hat\sigma_z(-\infty)\rangle=-1$, it is convenient to make the replacement  $\hat\sigma_z\to -1 $ straightaway in Eq.
(\ref{motion_1}). As a result we come to
\begin{eqnarray}
        \label{motion_1a}
        \frac{\partial}{\partial t}\langle\hat\sigma_-^{}\rangle=-i(\omega_0+\Delta\omega)\langle\hat\sigma_-^{}\rangle -
        i\int\limits_0^\infty g_\omega  \langle\hat{a}_\omega \rangle\,d\omega\,.
\end{eqnarray}
The integration of this equation with respect to time yields
\begin{eqnarray}
        \label{motion_1b}
        \langle\hat\sigma_-^{}\rangle=-i\int\limits_0^\infty\! g_\omega \! \int\limits_{-\infty}^t\!\!
        \langle\hat{a}_\omega(t')\rangle\, e^{-i(\omega_0+\Delta\omega)(t-t')} dt'\,d\omega\,.
\end{eqnarray}
The substitution of (\ref{motion_1b}) into (\ref{motion_1}) gives us 
\begin{eqnarray}
        \label{motion_1c}
        &&\frac{\partial}{\partial t}\hat\sigma_-^{}=-i\omega_0\hat\sigma_-^{}-
        i\int\limits_0^\infty g_\omega  \hat{\Theta}_\omega \,d\omega\,,
\end{eqnarray}
where
\begin{eqnarray}
 \label{motion_1d}
 \hat{\Theta}_\omega(t)=\hat{a}_\omega(t)+i\Delta\omega\!\!\int\limits_{-\infty}^t
    \!\!\langle\hat{a}_\omega(t')\rangle\, e^{-i(\omega_0+\Delta\omega)(t-t')} dt'\,.~~~
 \end{eqnarray}
By integration of (\ref{motion_1c}) we obtain the explicit relation 
\begin{eqnarray}
        \label{motion_1e}
        \hat\sigma_-^{}=-i\int\limits_0^\infty\! g_\omega \! \int\limits_{-\infty}^t\!\!
        \hat{\Theta}_\omega(t')\, e^{-i\omega_0(t-t')} dt'\,d\omega\,
\end{eqnarray}
between the operators $\hat{a}_\omega$ and $\hat{\sigma}_-^{}$, which defines the interaction of the QD--exciton with quantum light in the weak--field limit.  This equation serves as a basis for the further analysis because it
allows us to develop the correlation between the electric field and the polarization operators in QD. Going to the frequency domain and utilizing the relation
$\widehat{\mathbf{P}}(\mathbf{r})=|\xi(\mathbf{r})|^2{\bm\mu}^*\hat\sigma_-^{}+\hbox{H.c.}\,$ and Eqs. (\ref{E_0}), (\ref{motion_1d}) and (\ref{motion_1e}) we come to the frequency--domain polarizability operator as
\begin{equation}\label{a4}
    \widehat{\mathbf{P}}(\mathbf{r})=-\frac{|{\bm\mu}|^2/\hbar V}
    {\omega-\omega_0+i 0}
    \left[\widehat{\bf E}_0({\bf r})+
    \frac{\Delta\omega\,\langle \widehat{\bf E}_0({\bf r})\rangle}
    {\omega-\omega_0-\Delta\omega+i 0}\right]\,.
\end{equation}
with $\widehat{\bf E}_0({\bf r})$ given by Eq. (\ref{E_0}).  

Let us restrict further consideration to monochromatic fields.  This allows us to omit from here onwards the index
$\omega$ in operators  $\hat{a}_\omega$ and $\hat{a}_\omega^\dag$ and omit explicit mention of the frequency dependence in complex--amplitude operators.

\subsection{Forward scattering amplitude}

Let the field $\widehat{\bf E}_0({\bf r})$ be scattered by an isolated QD. Obviously, the total field $\widehat{\bf E}({\bf r})$ outside the QD is a superposition
of the incident and the scattered fields. In terms of the polarization induced in the QD, this field is expressed  by  the relation
\begin{eqnarray}
\label{a2}
  \widehat{\bf E}({\bf r})=\widehat{\bf E}_0({\bf r})+
    (\nabla\nabla\cdot +k^2)\!\!\int\limits_{V}
    \widehat{\bf P}({\bf r}')\,
    \frac{e^{ik|{\bf r}-{\bf r}'|}}{|{\bf r}-{\bf r}'|}d^3{\bf r}'\,,~~
\end{eqnarray}
which follows from the Maxwell equations. In turn, the induced polarization  is related to  a weak incident field by the linear relation (\ref{a4}).
Substitution of  (\ref{a4}) into (\ref{a2}) allows to couple  the incident and the total field, that is, to define the scattering operator \cite{Beresteckii}.

The analogous relation with operators  replaced by corresponding $C$-numbers holds for classical fields. For electrically small scatterers, the coupling
between the polarization and the electric field is local. This allows the introduction of the forward scattering amplitude by
\begin{eqnarray}
\label{b2}
  {\bf P}({\bf r}_0)=\frac{V}{k^2}f(0){\bf E}({\bf r}_0)\,,
\end{eqnarray}
where   ${\bf r}_0$ is the scatterer radius--vector. In the far zone, the scattered field  is a diverging spherical wave; thus,
\begin{eqnarray}\label{a1}
  {\bf E}({\bf r})={\bf E}_0({\bf r})+
    \frac{V}{k^2}(\nabla\nabla\cdot +k^2)
    \frac{e^{ik|{\bf r}-{\bf r}_0|}}{|{\bf r}-{\bf r}_0|}{f}(0){\bf E}_0({\bf r}_0) \,.~
\end{eqnarray}
Then,  the forward scattering amplitude ${f}(0)$ is found to be proportional to the QD polarizability \cite{Slepyan_pra02} and is given by 
\begin{equation}
\label{ampl}
f(0)= -\frac{k^2|{\bm\mu}|^2}{\hbar V}
\frac{1}{\omega-\omega_{0}-\Delta\omega+i 0}\,.
\end{equation}

Going to quantum light we cannot restrict ourselves to the formal replacement  of electric field and polarization in (\ref{b2}) and (\ref{a1}) by corresponding
operators, keeping forward scattering amplitude in the form of (\ref{ampl}). As follows from Eq. (\ref{a4}), because of the local field impact, the
polarization operator is represented by a superposition of the electric field operator and its mean value. As a result, Eq. (\ref{a4}) can not be reduced to
Eq. (\ref{b2}) written for operators.  

The necessary generalization of (\ref{b2}) exploits representation of the forward amplitude as an operator in the space of
quantum states of light:
\begin{eqnarray}\label{a1_1}
  \widehat{\bf P}({\bf r}_0)=
    \frac{V}{2k^2}\Bigl[\hat{f}(0)\widehat{\bf E}_0({\bf r}_0)+\widehat{\bf E}_0({\bf r}_0)\hat{f}(0)\Bigr]\,.
\end{eqnarray}
As one can see, the product of the operators $\hat{f}(0)$ and $\widehat{\bf E}_0$ enters this equation in symmetrized form, analogously to the interaction
Hamiltonian (\ref{H_I}). Using (\ref{E_0}), (\ref{a4}) and (\ref{a1_1}), we derive the equation
\begin{equation}
\label{zero_amp}
\hat{f}(0)\hat{a}+\hat{a}\hat{f}(0)=2f_2(0)\hat{a}+2\langle \hat{a}\rangle[f_1(0)-f_2(0)]\,,
\end{equation}
for the forward scattering amplitude operator $\hat{f}(0)$, where
\begin{equation}
\label{zero_amp_12}
f_{1,2}(0)= -\frac{k^2|{\bm\mu}|^2}{\hbar V}
\frac{1}{\omega-\omega_{1,2}+i 0}
\end{equation}
are the partial forward scattering amplitudes with
$\omega_1=\omega_0+\Delta\omega$ and $\omega_2=\omega_0$. To solve (\ref{zero_amp}) with respect to $\hat{f}(0)$, we introduce the operator
\begin{equation}
\label{koefficients}
    \hat{\xi}=\sum_{m=0}^{\infty}\!\frac{|2m+1\rangle\langle 2m|}
    {\sqrt{2m+1}}\,,
\end{equation}
where $|n\rangle$ stands for the  $n$-th Fock state. The operator satisfies the identity
\begin{equation}
    \label{ident}
    \hat{\xi}\hat{a}+
\hat{a}\hat{\xi}=1\,.
\end{equation}

We snow seek solution of (\ref{zero_amp}) in the form of $\hat{f}(0)=X\hat{\xi}+Y$. Substituting this
representation into Eq. (\ref{zero_amp}) and equating the terms of the same order in  $\hat{a}$, we  find coefficients $X$ and $Y$ and, thus, write the
forward scattering amplitude operator explicitly:
\begin{equation}
\label{solution}
    \hat{f}(0)=2\langle \hat{a}\rangle[f_1(0)-f_2(0)]\hat{\xi}+f_2(0)\,.
\end{equation}
The total electric field operator is expressed by (\ref{a2}) with the polarization operator defined by (\ref{a1_1}).

\section{Quantum optics of excitonic composite}

\subsection{Constitutive relations for field operators}

Consider a heterogeneous medium comprising a cubic lattice of identical spherical QDs embedded in a semiconducting host.  The QD radius and distance between
QDs are assumed to be much less than the wavelength in the host medium. The optical properties of such a composite medium are isotropic and can be described by an
effective index of refraction. Let the optical density of the composite be small enough to neglect the electromagnetic coupling of QDs, i.e., to neglect the
Mossotti--Clausius correction. Since the forward
scattering amplitude turns out to be an operator, the optical response of the composite is described, instead of (\ref{index}), by the index of refraction
\begin{equation}
    \label{refraction index}
\hat{n}^2=1+\frac{4\pi \rho}{k^2}\hat{f}(0)\,,
\end{equation}
which  is \emph{an operator in the space of quantum states of light}. Obviously, the same is relevant to the relative permittivity
$\hat{\varepsilon}=\hat{n}^2$.  Utilizing the well--known relation
$\widehat{\mathbf{D}}=\widehat{\mathbf{E}}+4\pi\widehat{\mathbf{P}}$ and (\ref{a1_1}),  we come to the \emph{operator constitutive relation}
\begin{equation}
\widehat{\mathbf{D}}= \frac{1}{2}\bigl(
    \hat{\varepsilon}
    \widehat{\mathbf{E}}+\widehat{\mathbf{E}}
    \hat{\varepsilon}\bigr)\,,
    \label{const_rel_2}
\end{equation}
with the effective operator permittivity
\begin{equation}\label{dielectr}
  \hat{\varepsilon}=2\langle \hat{a}\rangle\,\hat{\xi}\,n_1^2+
  (1-2\langle \hat{a}\,\rangle\hat{\xi})\,n_2^2\,,
\end{equation}
where $n_{1,2}^2=1+4\pi \rho {f_{1,2}(0)}/{k^2}\,$.
The constitutive relations
(\ref{const_rel_2}) and (\ref{dielectr}) satisfy the principle of
correspondence: in the classical limit
$\widehat{\mathbf{D}} \to \langle\widehat{\mathbf{D}}\rangle$ we
come  to $ \langle \widehat{\mathbf{D}}\rangle=n_1^2 \langle
\widehat{\mathbf{E}}\rangle$, i.e., the correct constitutive
relation for an excitonic composite exposed to classical light
\cite{Slepyan_NATO_03,Maksimenko_ENN}. Correspondingly, in that case the forward scattering amplitude is a $C$--number and is expressed by (\ref{ampl}). Even so, the local fields may result in physically observable effects, such as polarization--dependent splitting of the gain line \cite{Slepyan_99a,Maksim_00a} in asymmetric QDs.  The splitting was observed experimentally in \cite{Gammon_PRL_96,Stufler_PRB_05} and was referred to as a manifestation of the exchange interaction. Let us emphasize once more time that here we deal with the same effect presented in phenomenological and microscopic languages.

Because of (\ref{ident}), the operators $\hat{\varepsilon}$ (\ref{dielectr}) and $\hat{a}$ do not commutate  and, consequently, the symmetrized representation of
the constitutive relation (\ref{const_rel_2}) is of fundamental importance. Note that the operator $\hat{\varepsilon}$  is non-Hermitian even in the absence of
dissipation. Its physical meaning we discuss later on. Here we only stress that the  non-hermitian nature  does not result in any physical
contradiction since the operator $\hat{\varepsilon}$  does not represent any observable quantity. In contrast, the electric displacement operator
$\widehat{\bm{\mathcal{D}}}$, related to its Fourier amplitude $\widehat{\mathbf{D}}$ by the the procedure analogous to Eqs. (\ref{E_+}), is Hermitian as it
must be for observable quantities. The hermicity is due to the symmetrized notation on the right--side part of Eq. (\ref{const_rel_2}).

\subsection{Generalized plane waves}

Consider the propagation in the excitonic composite of free
(transverse) electromagnetic waves. We assume $n_{1,2}^2>0$; thus we
exclude from the analysis narrow forbidden zones between
longitudinal and transverse resonances in the vicinity of
frequencies $\omega_{1,2}$. The propagation is described by  the 1D frequency--domain operator Maxwell equations
\begin{equation}\label{Max}
  ik\,\widehat{\mathbf{H}}={\bf e}_z\times
  \frac{\partial\widehat{\mathbf{E}}}{\partial z}\,,\quad
  -\frac{ik}{2}(\hat{\varepsilon}\,\widehat{\mathbf{E}}+
  \widehat{\mathbf{E}}\,\hat{\varepsilon})=
  {\bf e}_z\times\frac{\partial\widehat{\mathbf{H}}}{\partial z}\,.
\end{equation}
This system can easily be transformed to a set of wave equations for the  operator $\widehat{\mathbf{E}}-\langle \widehat{\mathbf{E}}\rangle$ and the mean value
$\langle \widehat{\mathbf{E}}\rangle$:
\begin{eqnarray}
\label{wave1}
&&  \Bigl(\frac{\partial^2}{\partial z^2}+k^2 n_2^2\Bigr)\langle \widehat{\mathbf{E}}\rangle=0\,,
    \\ \rule{0in}{4ex}
    \label{wave2}
&&  \Bigl(\frac{\partial^2}{\partial z^2}+k^2 n_1^2\Bigr)(\widehat{\mathbf{E}}-\langle \widehat{\mathbf{E}}\rangle)=0\,.
\end{eqnarray}
The elementary solution of these equations  leads to the expansions
\begin{equation}\label{single-mode2}
  \widehat{\mathbf{E}}=i\sqrt{\frac{\hbar k}{A}}\,{\bf e}_y
  \left[
  \frac{e^{ikn_1z}}{\sqrt{n_1}}\,\langle \hat{a}\rangle\,+
  \frac{e^{ikn_2z}}{\sqrt{n_2}}\,(\hat{a}-\langle \hat{a}\rangle)
  \,\right],
\end{equation}
\begin{equation}\label{single-mode2H}
  \widehat{\mathbf{H}}=-i\sqrt{\frac{\hbar k}{A}}\,{\bf e}_x
  \left[
  {e^{ikn_1z}}{\sqrt{n_1}}\langle \hat{a}\rangle+
  {e^{ikn_2z}}{\sqrt{n_2}}(\hat{a}-\langle \hat{a}\rangle )
  \right],
\end{equation}
for the  mode traveling in the direction $z>0$. Eqs. (\ref{single-mode2}) and (\ref{single-mode2H}) display  the local field effect
in the excitonic composite: decomposition of a single mode into
coherent and incoherent parts with different refractive indexes. Thus, as
quantum light propagates through the excitonic composite, the light statistics is altered and spatial modulation of the
amplitude with the period
\begin{equation}
    \label{period} L_0=\frac{2\pi}{|n_{1}-n_{2}|k}
\end{equation}
 occurs. The period is
strongly dependent on  frequency due to the frequency dependence of partial  refractive indexes $n_{1,2}$ determined by Eq. (\ref{zero_amp_12}). In turn, the
alteration leads to transformation of the light quantum coherence of the second and higher degrees. To some extent, the situation is similar to that in
nuclear optics \cite{Baryshevskii}: the existence of two spin--dependent refractive indexes of the particle's wavefunction causes the spin rotation as the
particle travels through a polarized medium.

Physically, the result obtained can be understood  as follows. There are two mechanisms of quantum light scattering by a QD. The first one --
quasi--classical -- is due to the induction in the QD of the observable polarization and, therefore, it persists in the classical limit. The corresponding
frequency turns out to be shifted by the value $\Delta\omega$. This mechanism changes the QD state due to the local--field induced electron--hole correlations
and thus provides inelastic channel of the scattering. The first term in the right side of (\ref{single-mode2}) conforms to this channel. The aforementioned
non-hermicity of the operator $\hat{\varepsilon}$ even in the absence of dissipation is due to inelastic scattering. The second mechanism manifests itself
in the scattering of states of light having zero observable field (e.g., a single Fock state). Such states  do not induce observable polarization and,
consequently, any frequency shift. This mechanism has no classical analog and provides elastic channel of the scattering. The presence of the inelastic scattering
channel signifies the change of the scattered light's quantum state, i.e., the photon statistics dispersion.   Thus, the photon statistics dispersion and the
resulting phenomenon of the electromagnetic field spatial modulation are the direct manifestations of the fine structure of excitonic line, inherent in an
isolated QD \cite{Slepyan_pra02}.

Going to time--domain representation of operators (\ref{single-mode2}) and (\ref{single-mode2H}),  $\widehat{\bm{\mathcal{E}}}(\mathbf{r},t)$ and
$\widehat{\bm{\mathcal{H}}}(\mathbf{r},t)$, by the procedure defined by Eq. (\ref{E_+}), we can show that the condition
\begin{equation}
\label{conserv_mom}
    \int\limits_A\langle\widehat{\bm{\mathcal{E}}}(\mathbf{r},t)\times\widehat{\bm{\mathcal{H}}}(\mathbf{r},t)\rangle\mathbf{e}_z\,dx\,dy=\hbox{const}
\end{equation}
holds true for an arbitrary quantum state. Physically, this condition reflects the momentum conservation for fields determined by operators (\ref{single-mode2}) and (\ref{single-mode2H}): a mean value of the energy flux through an arbitrary closed surface is equal to zero for an arbitrary quantum state of light. That is, dissipation  is absent in the medium in spite of the non--hermicity of the relative permittivity operator $\hat{\varepsilon}$. The conservation low (\ref{conserv_mom}) testifies to physcal correctness of electromagnetic waves defined by the operators (\ref{single-mode2}) and (\ref{single-mode2H}).

\subsection{Input--output relations}

Consider a planar layer of the QD-based composite material of thickness $d$ located at $|z|<d /2$ exposed to a quantum field. The problem to be solved is to ascertain relations between operators of  incoming, reflected and transmitted waves. For that, we make use of 
 Eqs. (\ref{single-mode2}) and (\ref{single-mode2H}) and utilize the layer's reflection $R$ and transmission $T$ amplitudes (see, e.g., Eqs. (6.130)-(6.132) in Ref. \cite{Welsch_b01}). The reflection and transmission amplitudes satisfy the property $|R|^2+|T|^2=1$ which is a reflection of the scattering matrix's unitarity. Obviously, the total field outside the layer can be  represented  by \cite{Welsch_b01}
\begin{equation}
\label{outside}
\widehat{\mathbf{E}}=\widehat{\mathbf{E}}_{+}^\mathrm{in}+\widehat{\mathbf{E}}_{+}^\mathrm{out}+
    \widehat{\mathbf{E}}_{-}^\mathrm{in}+\widehat{\mathbf{E}}_{-}^\mathrm{out}\,,
\end{equation}
where $\widehat{\mathbf{E}}_{\pm}^\mathrm{in}$ are the operators of incoming waves propagating along the positive (index ''$+$'') and negative (index ''$-$'') directions of the axis $z$. The superscript ''out'' marks outgoing waves. By analogy with (\ref{E_0}) the incoming waves are given by
\begin{equation}
\label{E_in}
    \widehat{\mathbf{E}}_{\pm}^\mathrm{in}=i\sqrt{\frac{\hbar k}{A}}\,
    {\bf e}_y\,\hat{a}_{\pm}\,e^{\pm ikz}\,,
\end{equation}
where creation/annigilation operators $\hat{a}_{\pm}^\dag/\hat{a}_{\pm}$ satisfy ordinary bosonic commutation relations.

In view of the commutativity of operators  0n  the right sides of Eqs. (\ref{single-mode2})  and (\ref{single-mode2H}), the layer's transmission/reflection
balance  can be written separately for each of waves $\exp(ikn_{1,2}z)$:
\begin{equation}
    \label{bal_1}
\langle \hat{a}^\mathrm{out}_{\pm}\rangle=T_1\langle \hat{a}_{\pm}\rangle+ R_1\langle \hat{a}_{\mp}\rangle\,,
\end{equation}
\begin{equation}
        \label{bal_2}
\hat{a}^\mathrm{out}_{\pm}-\langle \hat{a}^\mathrm{out}_{\pm}\rangle
    =T_2(\hat{a}_{\pm}-\langle \hat{a}_{\pm}\rangle)+ R_2(\hat{a}_{\mp}-\langle \hat{a}_{\mp}\rangle)\,.
\end{equation}
Here $T_{1,2}$ and $R_{1,2}$ are the transmission and reflection coefficients of layers of the thickness $d$ and refractive indexes $n_1$ and $n_2$,
respectively. Using (\ref{bal_1})--(\ref{bal_2}), we can represent outgoing fields as
\begin{eqnarray}
        \widehat{\mathbf{E}}_{\pm}^\mathrm{out}&=&i\sqrt{\frac{\hbar k}{A}}\,
    {\bf e}_y\,e^{\pm ikz}
    \left[T_2\hat{a}_{\pm}+R_2\hat{a}_{\mp}\right.
    \cr \rule{0in}{4ex}
    &&\left. +(T_2-T_1)\langle \hat{a}_{\pm}\rangle+(R_1-R_2)\langle \hat{a}_{\mp}\rangle
    \right]\,.\label{E_out}
\end{eqnarray}
It can easily be shown that,  in view of $|R_{1,2}|^2+|T_{1,2}|^2=1$, the outgoing electromagnetic field satisfies the correct commutation relations for field
operators and the energy--momentum balance condition for incoming and outgoing waves for light in arbitrary quantum state.

Equations (\ref{E_out}) show that the quantum light transmission through  a layer of excitonic composite leads to the alteration of light coherence. For
example,  using the obvious identity $\langle(\hat{a}_{\pm}-\langle \hat{a}_{\pm}\rangle)^\dag (\hat{a}_{\pm}-\langle \hat{a}_{\pm}\rangle)\rangle=
\langle\hat{a}_{\pm}^\dag\hat{a}_{\pm}\rangle-|\langle\hat{a}_{\pm}\rangle|^2$, we derive
\begin{equation}
\label{photo}
\langle|\widehat{\mathbf{E}}_{+}^\mathrm{out}|^2\rangle \sim |T_2|^2\langle \hat{a}_{+}^\dag\hat{a}_{+}\rangle+
    (|T_1|^2-|T_2|^2)|\langle \hat{a}_{+}\rangle|^2\,.
\end{equation}
The photocurrent measured by a detector located in the region $z>d/2$ is proportional to $\langle|\widehat{\mathbf{E}}_{+}^\mathrm{out}|^2\rangle   $. Thus,
the photodetector signal is no longer completely determined by the mean number of photons (first term in (\ref{photo})) but depends also on the light statistics
(second term in (\ref{photo})). This effect can be exploited for experimental verification of theoretical predictions of the present article. The distinction
of the reflection/transmission properties of the excitonic composite layer for coherent and non-coherent components can be used for the design of new
quantum--optical light transformers.

\section{Photonic--state dispersion}

The complete characterization of a quantum state of light is given by the density operator.  In this section, we evaluate this operator in an excitonic composite
medium. Along with that, the issues of interest are the light characteristics measurable in typical photodetection processes. As an example of such
characteristics,  further we consider the normally--ordered normalized time--zero second--order correlation function $g^{(2)}(z)$ of the excitonic composite,
which characterizes the quantum--mechanical second--order degree of coherence at a given cross--section $z$ \cite{Scully_b01}.

\subsection{Density operator}

Let us rewrite Eq. (\ref{single-mode2})  as
\begin{equation}\label{single-mode3}
  \widehat{\mathbf{E}}=i\sqrt{\frac{\hbar k}{An_2}}\,{e^{ikn_2z}}\hat{\alpha}(z)\,{\bf e}_y\,,
  \end{equation}
where $\hat{\alpha}(z)=\hat{a}+\kappa(z)$ and
\begin{equation}\label{kappa_0}
\kappa(z)=\langle\hat{a}\rangle\left[\sqrt{\frac{n_2}{n_1}}e^{i(n_1-n_2)kz}-1\right]\,.
\end{equation}
The quantity $\kappa(z)$ is a periodic function of the spatial coordinate $z$ with oscillation period $L_0$ (\ref{period}).  Operators $\hat{\alpha}(z)$
and $\hat{\alpha}^\dag(z)$ satisfy the ordinary bosonic commutation relations. That is, they can be treated as annihilation and
creation operators of the quantums of some bosonic field. Utilizing then operator identities (3.47) and (3.48) from \cite{Welsch_b01}, we come to
\begin{eqnarray}
\label{alpha_2}
    &&\hat{\alpha}=\hat{D}_\kappa^\dag\hat{a}\hat{D}_\kappa\,,\\
    \noalign{\hbox{where}}
    &&\hat{D}_\kappa=\exp(\kappa\hat{a}^\dag-\kappa^*\hat{a})
\end{eqnarray}
is the displacement operator with $\kappa$ as the displacement factor.  Hereafter we omit the $z$--dependence. According to (\ref{alpha_2}), as
light goes through the excitonic composite the spatial evolution of the light density operator  is governed  by the unitary transform
\begin{equation}
    \label{dens_oper}
\hat{\varrho}'=\hat{D}_\kappa^\dag\hat{\varrho}\hat{D}_\kappa\,,
\end{equation}
where
$\hat{\varrho}$ is the initial operator of  the light density induced by a source at the input.

The distinguishing property of the transform (\ref{dens_oper}) is that it actually is \emph{not linear}: the displacement factor $\kappa$ depends on the state of
initial quantum light through the relation
\begin{equation}
    \label{kappa}
\kappa\sim \langle\hat{a}\rangle=\sum_{n=0}^\infty \sqrt{n+1}\,\varrho_{n+1,n}\,.
\end{equation}
Physically, this means that the field acting on the QD--exciton is self--consistent in the  Hartree--Fock--Bogoliubov sense  (see Eqs. (\ref{Delta_Ha}) and
(\ref{Delta_H})). As a result of self--consistency, nonlinearity of the quantum--mechanical motion of carriers in QD arises;  see the second term on the
right side of Eq. (\ref{motion_1}).

The operator transformation law (\ref{dens_oper})  can be rewritten in terms of familiar matrix elements. As the first step we represent $\hat{\varrho}'$ by the
expansion
\begin{equation}
    \label{dens_oper-1}
\hat{\varrho}'=\sum_{m,n}\varrho_{mn}\hat{D}_{-\kappa}|m\rangle\langle n|\hat{D}_{-\kappa}^\dag\,.
\end{equation}
In accordance with Ref. \cite{Welsch_b01},
\begin{equation}
    \label{dens_oper-2}
\hat{D}_{-\kappa}|m\rangle=\exp\Bigl(-\frac{|\kappa|^2}{2}\Bigr)\sum_{p}S_{mp}(\kappa)|p\rangle \,,
\end{equation}
where
\begin{equation}\label{kappa_2}
S_{mp}(\kappa)=\left\{
\begin{array}{ll}
(\kappa^*)^{m-p}\sqrt{\displaystyle\frac{p!}{m!}}\, L_p^{(m-p)}(|\kappa|^2)\,, & m \geq p\,, \\\rule{0in}{5ex}
(-\kappa)^{p-m} \sqrt{\displaystyle\frac{m!}{p!}} \,L_p^{(p-m)}(|\kappa|^2)\,, & m < p\,,
\end{array}\right.
\end{equation}
and $L_m^{(n)}(x)$ are the generalized Laguerre polynomials. Then, using (\ref{dens_oper-1})--(\ref{kappa_2}) we can proceed from (\ref{dens_oper}) to the law
of transformation of matrix elements:
\begin{equation}\label{tranform}
    \varrho_{pq}'= \exp\Bigl(-\frac{|\kappa|^2}{2}\Bigr)
 \sum_{m,n}\varrho_{mn}S_{mp}(\kappa)S_{nq}^*(\kappa)
 \,.
\end{equation}

Equations (\ref{dens_oper}) and (\ref{tranform}) determine the transformation  law for quantum light with arbitrary statistics.  Consider several particular
cases as examples. For a single Fock state $|n\rangle$, $\langle\hat{a}\rangle=0$ and $\kappa=0$. Taking $\hat{D}_{0}=1$, into account we conclude that the
excitonic composite does not transform single Fock states. Analogous  situation occurs for thermal light where $\varrho_{mn}\sim\delta_{mn}$ and, accordingly
(\ref{kappa}), $\langle\hat{a}\rangle=0$. The spatial evolution of a coherent state $|\beta\rangle$ is characterized by the transform 
$\varrho'=\hat{D}_{\kappa}^\dag|\beta\rangle\langle \beta|\hat{D}_{\kappa}=\hat{D}_{-\kappa}\hat{D}_{-\beta}|0\rangle\langle 0|\hat{D}_{-\beta}^\dag\hat{D}_{-\kappa}^\dag\,$. Taking into account the well--known identity \cite{Welsch_b01} $ \hat{D}_{-\kappa}\hat{D}_{-\beta}=\hat{D}_{-\beta-\kappa}\exp[i\mathrm{Im}(\kappa\beta^*)]$, we come to
\begin{eqnarray}
\label{tranform2}
\varrho'&=&|\kappa+\beta\rangle\langle\kappa+\beta| \cr\rule{0in}{4ex}
&=& \Bigl|\sqrt{\frac{n_2}{n_1}}\beta\,e^{i(n_1-n_2)kz}\Bigr\rangle\Bigl\langle\sqrt{\frac{n_2}{n_1}}\beta\,e^{i(n_1-n_2)kz}\Bigr|\,.	
\end{eqnarray}
This result implies that initial coherent state is transformed into another coherent state with modified coherent amplitude. The coherent amplitude of transformed state  experiences spatial beating as light passes the composite (factor $\exp[{i(n_1-n_2)kz}]\,$ in (\ref{tranform2})). The mean photon number  in the composite $\sqrt{{n_2}/{n_1}}|\beta|^2$ holds invariable and significantly differs from that the source produces in vacuum, $|\beta|^2$. The alteration is determined by the ratio $n_2/n_1$ which varies in  a broad range depending on the photon energy. There are different energy bands in the excitonic composite, where the classical source efficiency  either rises ($n_2>n_1$) or falls ($n_2<n_1$), in  comparison with the source efficiency in vacuum.  

As one more example, consider the transformation of the Fock qubit. Let $|\Psi\rangle=\beta_0 |0\rangle+\beta_1|1\rangle\,$, where the coefficients $\beta_{0,1}$ are coupled by the ordinary normalization $|\beta_{0}|^2+|\beta_{1}|^2=1\,$. According to (\ref{dens_oper}) and (\ref{kappa}), the transformation is given by 
\begin{eqnarray}
\label{tranform2a}
\varrho'&=&|\Psi'\rangle\langle\Psi'| 
	=\hat{D}_{-\kappa}|\Psi\rangle\langle\Psi|\hat{D}_{-\kappa}^\dag\,,	
\end{eqnarray}
with $\langle\hat{a}\rangle=\beta_{0}^*\beta_{1}$ in (\ref{kappa_0}) for the displacement factor $\kappa$.   Obviously, in the coherent--state basis  $|\Psi'\rangle=(\beta_0\hat{D}_{-\kappa}+\beta_1\hat{D}_{-\kappa}\hat{a}^\dag)|0\rangle$. Taking then into account the relation $[\hat{a}^\dag,\hat{D}_{-\kappa}]=-\partial \hat{D}_{-\kappa}/\partial\hat{a}=-\kappa^* \hat{D}_{-\kappa}$ (see [\onlinecite{Welsch_b01}, Eq. (C.17)]), we obtain 
$|\Psi'\rangle=(\beta_0+\beta_1\kappa^*+\beta_1\hat{a}^\dag)|\kappa\rangle\,$. Then, proceeding to the Fock--state representation
we come to the expression
\begin{eqnarray}
\label{tranform4}
|\Psi'\rangle=\exp\Bigl(-\frac{|\kappa|^2}{2}\Bigr)\!\!\sum_{n=0}^\infty\frac{\kappa^n}{\sqrt{n!}}
	\left(\beta_0+\beta_1\kappa^*+\beta_1\frac{n}{\kappa}\right)|n\rangle.~	
\end{eqnarray}
Equations (\ref{tranform2a}) and (\ref{tranform4}) define the spatial evolution of the Fock qubit in excitonic composite. Let us stress that (\ref{tranform4}) includes 
a full set of Fock states. An essential physical conclusion can be deduced from that: new states absent in light in the beginning arise as light travels through the excitonic composite. 

\subsection{Second--order coherence}

To illustrate the effect of the photon statistics  dispersion,  we
calculate the second--order correlation function $g^{(2)}(z)$, which    is
represented in terms of the field operators by \cite{Scully_b01}
$$g^{(2)}(z) =
\langle \widehat{\mathbf{E}}^\dag\widehat{\mathbf{E}}^\dag
\widehat{\mathbf{E}}\widehat{\mathbf{E}}\rangle/\langle
\widehat{\mathbf{E}}^\dag\widehat{\mathbf{E}}\rangle^2\,.
$$
Let us consider light prepared in an arbitrary quantum state $|\Psi\rangle$. Then, after some algebraic manipulations, we obtain the second--order correlation function
for the field operator (\ref{single-mode2})
\begin{widetext}
\begin{equation}\label{corrfunct}
g^{(2)}(z)=\frac{\sum_n |\kappa^2(z) A_{n-1}+
    2\kappa(z)\sqrt{n}A_{n}+
    \sqrt{n(n+1)}A_{n+1}|^2}
    {\left[\sum_n {n}|A_{n}|^2+\Bigl(\displaystyle\frac{n_2}{n_1}-1\Bigr)|\langle \hat{a}\rangle|^2 \right]^2}
\end{equation}
\end{widetext}
where $A_n=\langle n|\Psi\rangle$. Without taking local fields into account, the
correlation function $g^{(2)}(z)$ coincides with the correlation
function  $g^{(2)}_0=\sum_n {n(n+1)}|A_{n+1}|^2/ {\Bigl(\sum_n
{n}|A_{n}|^2\Bigr)^2}$ of the incident field
$\widehat{\bf E}_0$.  The distinguishing property of $g^{(2)}(z)$  is
its frequency dependence which is due to the pronounced frequency
dispersion $n_{1,2}=n_{1,2}(\omega)$ of the excitonic composite.
Also, the second--order correlation function  experiences the spatial
modulation with period $L_0$.

The most intriguing peculiarity in the photon statistics appears in
the vicinity of the resonance $\omega=\omega_2$. For the light
states with a large enough $\langle\hat{a}\rangle$, the inequality
$n_2|\langle\hat{a}\rangle|^2\gg 1$ holds true because of the
 growth of $n_2$ in the vicinity of the resonance. As follows from
(\ref{corrfunct}), in that case the spatial modulation becomes a
high--order infinitesimal and $g^{(2)}(z)\to 1$ independently of
the light state $|\Psi\rangle$; that is, the photon statistics tends to Poissonian (completely uncorrelated photons) as light propagates through the excitonic composite. This means that the excitonic composite reduces the photon correlation degree for photons with both positive ($g^{(2)}_0> 1$) and negative ($g^{(2)}_0< 1$) correlation. Indeed,  $g^{(2)}(z)<g^{(2)}_0$ in the  former case, whereas $g^{(2)}(z)>g^{(2)}_0$ in the last one.
Another situation arises for a pure Fock state
$|n\rangle$ with $\langle\hat{a}\rangle=0$. In this case Eq.
(\ref{corrfunct}) gives $g^{(2)}= 1 - 1/n$. That is, the
correlation function (\ref{corrfunct})
coincides with the free--space correlation function. In general,
negligible changing of the correlation function takes place for
all quasi-Fock states characterized by the condition
$n_2|\langle\hat{a}\rangle|^2\ll 1$.

A spatial modulation of the  second--order correlation  function of light is characteristic for the Hanbury--Brown--Twiss (HBT) interference \cite{Scully_b01}
of quantum light emitted by two sources. As different from the HBT--effect, the spatial modulation $g^{(2)}(z)$ (\ref{corrfunct}) in the excitonic composite
arises as a result of the interference of coherent and incoherent components of the field.   The transformation of the statistics of quantum light interacting with
a finite--thickness dielectric slab with a $C$--number relative permittivity was considered in Ref. \cite{Welsch_b01}.  The statistics transformation is provided by
the light reflection/refraction at boundaries, similar to that as a dispersionless dielectric slab modifies the frequency spectrum of a signal due to the
interference of waves reflected from boundaries. Another effect of such a type has been predicted in Ref. \cite{Lodahl_05}: appearance of spatial quantum
correlation in light propagating through a multiple scattering random medium. The quantum correlations depend on the state of quantum light and  are due to the
cross--correlation between different output modes. In our case, the  physical mechanism of the transformation is completely different: we predict the transformation
effect to be inherent to a homogeneous medium. That allows us to identify the effect as \emph{a new type of optical dispersion}.

It should be noted that the notions ''linear regime'' and ''weak coupling regime'' as applied to excitonic composite mediums are not identical. In the presence of
local field, the polarization (\ref{a4}) is linear in the incident field (linear regime) but contains a quadratic term in the oscillator strength, i.e., the term
$O(|\bm{\mu}|^4)$.  Nonlinearity of that type violates the weak coupling regime and  leads to the transformation of the photon statistics similar to that takes place in
the standard Jaynes--Gummings dynamics \cite{Scully_b01}. In other words, in the presence of local field the light--QD interaction is characterized by two
coupling  parameters: the standard Rabi frequency and the new depolarization shift $\Delta\omega$.

\section{Summary and outlook}

In the present paper  we have presented a theory of the quantum light interaction with excitonic composites whose relative permittivity is predicted to be an operator in the space of quantum states of light. As a result, the medium's refractive index turns out to be dependent on the photon number distribution providing, by analogy with the frequency dispersion and the spatial dispersion,  \emph{the photon statistics
dispersion} of light.   Self-organized lattices of
ordered QD-molecules and 1D--ordered (In,Ga)As QD--arrays
\cite{Mano_04,Lippen_04} serve as examples of high--quality excitonic
composites.

Light transformation in the excitonic composite is carried out by means of the displacement operator $\hat{D}_\kappa$ --- one of the canonical transforms of quantum optics \cite{Scully_b01}. However, as different from ordinary cases, the displacement factor $\kappa$ in an excitonic composite depends not only on the photon energy and momentum (in mediums with spatial dispersion) but also on the photon statistics.  In particular, an excitonic composite provides \emph{different propagation regimes for coherent and non-coherent components of total field}. That is, the frequency response characteristics of excitonic composite--based structures may differ in  coherent and non-coherent light. For example, a distributed Bragg reflector with layered components made of excitonic composite materials would manifest shifted reflection bands for coherent and non-coherent light. In the same manner, an excitonic composite--based microcavity will possess, depending on the light statistics, two alternate sets of eigenfrequencies. An interferometer constituted by two collinear plane
mirrors of excitonic composite material can serve as the simplest model. The excitonic composite provide a new (not related to anisotropy) birefringence mechanism: coherent and non-coherent components of a plane wave obliquely incident on plane excitonic composite interface will propagate at different angles. 
Equations (\ref{bal_1}) and (\ref{bal_2}) shows that an excitonic composite layer can serve as a beam splitter with a modified law of transformation of the field operators. 
Along with the second--order coherence transformation, the optical effects mentioned create a potentiality for experimental verification of the predicted photon--statistics dispersion.

The theory elaborated in the present paper deals with the single--mode case that implies the single pair of creation/annihilation operators sufficient to characterize the fields (\ref{single-mode2}) and (\ref{single-mode2H}). In the multi--mode case each mode will be characterized by its own mode--number--dependent value $\langle\hat{a}\rangle $ and, correspondingly, permittivity operator. As a result, then the photon statistics dispersion will be accompanied by spatial dispersion. In classical crystal optics \cite{Agranovich_b84}, spatial dispersion necessarily entails optical anisotropy even in structurally isotropic mediums. 
The preferential direction is determined by the propagating wavevector and refractive indexes are different for transverse and longitudinal field components. In an isotropic medium  with the multi--mode photon--statistics dispersion we meet another situation: in spite of the spatial nonlocality the permittivity operator is scalar. The situation can be understood from the following.  A preferential direction is absent in an excitonic composite since it can be related only  to a physically observable wave. In contrast, in the case considered the spatial dispersion mechanism concerns with operators and disappears on proceeding to observable fields.

The excitonic composite considered in this paper is an example that demonstrates the existence of mediums characterized by the operator constitutive relation (\ref{const_rel_2}) and, consequently, displaying the photon statistics dispersion (\ref{dielectr}). It would be of interest to search other artificial (and, possibly, natural) materials with photon statistics dispersion and the statistics transformation law not necessarily defined by (\ref{dielectr}) and (\ref{dens_oper}), respectively. 
Since the photonic state dispersion in excitonic composites is due to dipole--dipole interactions in QDs,  one can  expect  the dispersion of that type in  mediums with strong dipole--dipole interaction, such as Bose--Einstein condensates
\cite{Lewenstein_94} and ultracold atomic ensembles \cite{Zhang_94}.

The developed model of quantum--optical properties of excitonic composites leads to a set of functional equations, in which both quantum fields to be found and predetermined coefficients are operators in the same space of quantum states. The investigation of general properties and methods of solution of such equations constitutes an intriguing mathematical problem. One can expect that the equations of that type will find application in studying of quantum fields of different physical origin.  The matter of special interest would be the case  when the equation solution and the coefficients are non-commutative --- as it takes place in our paper.

\acknowledgments
The work of S. A. Maksimenko was carried out during a stay
at the Institute for Solid State Physics, TU Berlin, and was supported
by the Deutsche Forschungsgemeinschaft (DFG). Authors acknowledge a support from  INTAS under project 
No 05-1000008-7801.

\appendix

\section{Envelope method for the local field--induced contribution to the Hamiltonian}

The application of the envelope representation (\ref{3:quantum_fields}) to the Hamiltonian $\cal{H}_{\rm C}$ given by Eqs. (\ref{HC_1}) and (\ref{H_C2}) makes possible averaging of (\ref{HC_1}) and (\ref{H_C2}) over unit cell and passing to continuous field of envelopes, i.e., allows to rewrite $\cal{H}_{\rm C}$  in terms of (\ref{HC_0})  and (\ref{Delta_Ha}). In spite of the fact that the procedure is well known (see, e.g., \cite{Yu_b01}), its realization in our case contains some peculiarities and thus requires a more detailed presentation, which  is given for the component $\cal{H}_{\rm C1}$ in the following. The case of $\cal{H}_{\rm C2}$ can be considered by analogy.

First, we stress that Eqs. (\ref{13:commut}) and (\ref{3:quantum_fields}) produce the commutation relations for envelopes:
\begin{equation}
    \begin{array}{lcl}
    [\hat{\xi}^\dag_n(\mathbf{R}_{q}),
    \hat {\xi}_{n'}(\mathbf{R}_{q'})]_{+}
    &=&
    \delta_{nn'}\delta_{qq'} \,,
    \\ \rule{0in}{5ex}
    [\hat{\xi}_{n}(\mathbf{R}_{q}),\hat {\xi}_{n'}(\mathbf{R}_{q'})]_{+}
    &=& [\hat {\xi}^\dag_{n}(\mathbf{R}_{q}),\hat
    {\xi}^\dag_{n'}(\mathbf{R}_{q'})]_{+} = 0\,.
    \end{array}
    \label{A:anticommut}
    \end{equation}
Substitution of  (\ref{3:quantum_fields}) into   (\ref{HC_1}) leads to
\begin{widetext}
\begin{eqnarray}
\label{A:eq2}
&&{\cal H}_\mathrm{C1}^{}= \left(\frac{eV}{N}\right)^2
\sum_{\substack{m,m' \\ n,n'}}\sum_{\substack{p,p' \\ q,q'}}
    \hat\xi_n^\dag(\mathbf{R}_p)\,\hat{\xi}_{n'}(\mathbf{R}_{p'})\,
    \langle \hat\xi_m^\dag(\mathbf{R}_q)\hat {\xi}_{m'}(\mathbf{R}_{q'})\rangle
    K^{pp'qq'}_{nn'mm'} \,,
\end{eqnarray}    
where
\begin{eqnarray}
\label{A:_eq3}
&&    K^{pp'qq'}_{nn'mm'} = N^2\int\limits_{V_p}\!\!\int\limits_{V_q}
    \frac
    {w_{np}^*(\mathbf{r})w_{n'p'}(\mathbf{r})w_{mq}^*(\mathbf{r}')w_{m'q'}(\mathbf{r}')}
    {|\mathbf{r}-\mathbf{r}'+\mathbf{R}_p-\mathbf{R}_q|}
        \,{d^3\mathbf{r}\,d^3\mathbf{r}'}\,.
\end{eqnarray}
\end{widetext}
and $V_p$ is the volume of the $p$-th unit cell.

For evaluation of integral (\ref{A:_eq3}) we introduce vectors $\mathbf{x}=\mathbf{R}_p-\mathbf{R}_q$ and $ \mathbf{y} =\mathbf{r}-\mathbf{r}'$ and take into account that $ \mathbf{y}\ll \mathbf{x}$ within the unit cell. This allows us to represent the denominator of (\ref{A:_eq3}) by the expansion
\begin{equation}
\label{A:_eq4}
    \frac{1}{|\mathbf{x}+\mathbf{y}|}\approx \frac{1}{|\mathbf{x}|}+ \mathbf{y}\cdot\nabla_\mathbf{x}\frac{1}{|\mathbf{x}|}+\frac{1}{2}
    \mathbf{y}\cdot\left(\nabla_\mathbf{x}\otimes\nabla_\mathbf{x}\frac{1}{|\mathbf{x}|}\right)\mathbf{y}+\dots,
\end{equation}
In the standard envelope technique, the third term is  supposed to be a small correction to the Coulomb interaction and conventionally neglected \cite{Yu_b01}. But, in our paper we demonstrate that the correction can be of importance and may result in qualitatively new effects.

Substitution of (\ref{A:_eq3}) into (\ref{A:_eq4}) leads us, in view of the orthonormality of the  Wannier  functions $w_{np}(\mathbf{r})$, to
\begin{eqnarray}
\label{A:eq5}
&&K^{pp'qq'}_{nn'mm'} =\frac{\delta_{nn'}\delta_{mm'}\delta_{pp'}\delta_{qq'}}{|\mathbf{R}_p-\mathbf{R}_q|}
    +\frac{N^2}{2}\mathbf{G}_{nn'pp'} \cdot \cr \rule{0in}{4ex}
    &&\qquad \left(\nabla_\mathbf{x}\otimes\nabla_\mathbf{x}\frac{1}{|\mathbf{R}_p-\mathbf{R}_q|}\right)\mathbf{G}_{mm'qq'}
     \,,\\
        \noalign{\hbox{where}} \rule{0in}{4ex}
\label{A:eq6}
&&     \mathbf{G}_{nn'pp'}=\int\limits_{V_p}
    \mathbf{r} w_{np}^*(\mathbf{r})w_{n'p'}(\mathbf{r})
        \,d^3\mathbf{r}\,.
\end{eqnarray}
For further consideration it is convenient to  turn from Wannier  to Bloch functions by means of the procedure
\begin{equation}
\label{A:eq7}
w_{np}(\mathbf{r})=\frac{1}{\sqrt{N}}\sum_k\exp(i\mathbf{k}\mathbf{R}_p)u_{nk}(\mathbf{r})\,.
\end{equation}
Using this representation we reduce Eq. (\ref{A:eq6}) to
\begin{eqnarray}
\label{A:eq8}
\mathbf{G}_{nn'pp'}=\frac{V_p}{eN}\sum_{k,k'}\exp[i(\mathbf{k}'\mathbf{R}_{p'}-\mathbf{k}\mathbf{R}_{p})]\bm{\mu}_{nn'kk'}\,,
 \end{eqnarray}
where
\begin{eqnarray}\label{A:eq9}
 \bm{\mu}_{nn'kk'}=eN\int\limits_{V_p}\mathbf{r}u_{nk}^*(\mathbf{r})u_{n'k'}(\mathbf{r})\,d^3\mathbf{r}\,
\end{eqnarray}
is the matrix element of the dipole transition in a 3D medium the same as the QD material. As a rule, the most important contribution into dipole moment is from direct transitions in the vicinity of the bottom of band \cite{Yu_b01}. For such transitions we can neglect the $k,k'$ dependences in the last equation and take $\bm{\mu}_{nn'kk'}=\bm{\mu}_{nn'}$ thereafter. Then, taking the identity $\sum_k\exp[i\mathbf{k}(\mathbf{R}_{p'}-\mathbf{R}_{p})]=N\delta_{pp'}$, we see that Eq.  (\ref{A:eq6}) can easily be reduced to
\begin{eqnarray}
\label{A:eq11}
 &&   \mathbf{G}_{nn'pp'}=\frac{1}{eN}\delta_{pp'}\bm{\mu}_{nn'}\,.
\end{eqnarray}
The concluding step is to turn to the continuous field of envelopes. For that we substitute (\ref{A:eq5})
 and (\ref{A:eq11}) into (\ref{A:eq2}) and perform the replacement $\mathbf{R}_{p'}\to \mathbf{r}$ and $\mathbf{R}_{q}\to \mathbf{r}'$ to get
\begin{equation}
    \begin{array}{l}
    \frac{\displaystyle V}{\displaystyle N}\sum_p\to \displaystyle \int d^3\mathbf{r}\,,
    \\ \rule{0in}{5ex}
    \nabla_{{\bf R}_p}\otimes\nabla_{{\bf R}_p}\frac{\displaystyle 1}{\displaystyle|{\bf R}_p-{\bf R}_q|} \to \underline{G}({\bf r}-{\bf r}') \,.
    \end{array}
    \label{A:eq12}
    \end{equation}
\\


\begin{thebibliography}{99}

\bibitem{Scully_b01} M. O. Scully and  M. S. Zubairy,
\textit{Quantum Optics} (University Press, Cambridge, 2001).

\bibitem{Knill_Nature_01} E. Knill, R. Laflamme and G. J. Milburn, Nature (London) {\bf 409}, 46 (2001).

\bibitem{Sanaka_PRL_04} K. Sanaka, T. Jennewein, J.-W. Pan, K. Resch and A. Zeilinger, Phys. Rev. Lett. {\bf 92}, 017902 (2004).

\bibitem{Sanaka_PRL_06} K. Sanaka, K. Resch and A. Zeilinger, Phys. Rev. Lett. {\bf 96}, 083601 (2006).

\bibitem{Lu_02}Y. J. Lu and Z. Y. Ou, Phys. Rev. Lett. {\bf 88}, 023601 (2002).

\bibitem{Lounis_05} B. Lounis and M. Orrit, Rep. Prog. Phys.  {\bf 68}, 1129 (2005).

\bibitem{Short_PRL_83} R. Short, and L. Mandel, Phys. Rev. Lett. {\bf 51}, 384 (1983).

\bibitem{Keever_Nature_03} J. McKeever, A. Boca, A. D. Boozer, J. R. Buck and H. J. Kimble, Nature (London) {\bf 425}, 268 (2003).

\bibitem{Choi_PRL_06} W. Choi, J.-H. Lee, K. An, C. Fang--Yen, R. R. Dasari and M. S. Feld, Phys. Rev. Lett. {\bf 96}, 093603 (2006).

\bibitem{Law_PRL_96} C. K. Law and J. H. Eberly, Phys. Rev. Lett. {\bf 76}, 1055 (1996).



\bibitem{Slepyan_NATO_03} G. Ya. Slepyan, S. A. Maksimenko, A. Hoffmann, and  D. Bimberg,  in \textit{Advances in Electromagnetics
of Complex Media and Metamaterials}, edited by S. Zouhdi
\textit{et al.} (Kluwer, Dordrecht, 2003), pp. 385-402.

\bibitem{Maksimenko_ENN} S. A. Maksimenko and G. Ya. Slepyan,
in \textit{Encyclopedia of Nanoscience and Nanotechnology,} edited
by J.A. Schwarz \textit{et al.}, (Marcel Dekker, New York, 2004),
pp. 3097 - 3107.

\bibitem{bimberg_b99} D. Bimberg, M. Grundmann, and
N. N. Ledentsov, {\it Quantum Dot Hetero\-structures} (Wiley,
Chichester, 1999).

\bibitem{Yamamoto_b} Y. Yamamoto, F. Tassone, and H. Cao,
\textit{Semiconductor Cavity Quantum Electrodynamics} (Springer
Verlag, Heidelberg, 2000)

\bibitem{Michler_b03} \textit{Single Quantum Dots, Topics of Applied Physics,}
edited by P. Michler (Springer Verlag, Heidelberg, 2003).

\bibitem{Michler_00} P. Michler, A. Imamoglu, M. D. Mason, P. J. Carson, G. F.
Strouse, and S. K. Buratto, Nature (London) {\bf 406}, 968 (2000).

\bibitem{Santori_PRL_01} C. Santori,  M.  Pelton, G.  Solomon, Y. Dale, and Y.  Yamamoto,  Phys. Rev. Lett. {\bf 86}, 1502
(2001). 

\bibitem{Moreau_PRL_02} E. Moreau, I. Robert, L. Manin, V. Thierry-Mieg, J. M. G\'erard and I. Abram, Phys. Rev. Lett. {\bf 87}, 183601 (2002).

\bibitem{Regelman_01} D. V. Regelman, U. Mizrahi, D. Gershoni, E. Ehrenfreund, W. V. Schoenfeld and P. M. Petroff, Phys. Rev. Lett. {\bf 87}, 257401 (2002).

\bibitem{Gerardot_PRL_05}B. D. Gerardot, S. Straut, M. J. A. de Dood, A. V. Bychkov, A. Bodoloto, K. Hennessy, E. L. Hu, D. Bouwmeester and P. M. Petroff, Phys. Rev. Lett. {\bf 95}, 137403 (2005).

\bibitem{Slepyan_pra02}  G. Ya. Slepyan, S. A. Maksimenko, A. Hoffmann, and  D. Bimberg,
\pra {\bf 66}, 063804, (2002).

\bibitem{Maksimenko_HN04} S.A. Maksimenko and G.Ya. Slepyan, in
\textit{The Handbook of Nanotechnology: Nanometer Structure Theory,
Modeling, and Simulation,} edited by A. Lakhtakia (SPIE Press, Belingham, 2004), pp. 145-206.

\bibitem{Ajiki_02} H. Ajiki, T. Tsuji, K. Kawano, and K. Cho,
Phys. Rev. B {\bf 66}, 245322 (2002).

\bibitem{Goupalov_03} S. V. Goupalov, Phys. Rev. B {\bf 68}, 125311 (2003).

\bibitem{Lewenstein_94} M. Lewenstein, Li You, J. Cooper, and K. Burnett, \pra \textbf{50}, 2207 (1994).


\bibitem{Welsch_b01}  W. Vogel, D.-G. Welsch and S. Wallentowitz,
{\it Quantum Optics. An Introduction} (Willey-VCH, New York,
2001).

\bibitem{Goldberger} M. L. Goldberger and K. M. Watson, {\it Collision Theory}, (Wiley, New York, 1964).

\bibitem{Landau_QM}  L.D.~Landau and E.M.~Lifshitz,
\textit{Quantum Mechanics}, 3rd edn. (Pergamon Press, Oxford, 1977).

\bibitem{Baryshevskii} V. G. Baryshevsky, Phys. Lett. A {\bf 171}, 431 (1992).
See also V.G. Baryshevsky, \textit{Nuclear Optics of Polarized Media}
(Energoatomizdat, Moscow, 1995, in Russian).

\bibitem{Cho_b03} K. Cho, \textit{Optical response in nanostructures}, (Springer Series in Solid-State
Science, Vol. 139, 2003).

\bibitem{Hang_b94} H. Haug and S. W. Koch, \textit{Quantum Theory of the Optical and Electronic Properties of Semiconductors} (World Scientific, Singapore, 1994).

\bibitem{Slepyan_99a}  G. Ya. Slepyan, S. A. Maksimenko, V. P. Kalosha, N. N.
Ledentsov, D. Bimberg and Zh. I. Alferov, Phys. Rev. B {\bf 59}, 1275
(1999).

\bibitem{Maksim_00a}  S. A. Maksimenko, G. Ya. Slepyan, V. P. Kalosha, S. V.
Maly, N. N. Ledentsov, J. Herrmann, A. Hoffmann, D. Bimberg, and Zh. I.
Alferov, J. Electronic Materials, \textbf{29}, 494 (2000).



\bibitem{Fleichhauer_pra_99} M. Fleichhauer and S. F. Yelin, \pra \textbf{59}, 2427 (1999).

\bibitem{Beresteckii} V. B. Beresteckii, E. M. Lifshitz, and L. P. Pitaevskii, {\it Quantum Electrodynamics} (Pergamon Press, Oxford, 1982).

\bibitem{Gammon_PRL_96}  D. Gammon, E. S. Snow, B. V. Shanabrook, D. S. Katzer, Phys. Rev. Lett. {\bf 76}, 3005 (1996)
and D. Park, Phys. Rev. Lett. {\bf 76}, 3005 (1996); Science {\bf 273}, 87 (1996).

\bibitem{Stufler_PRB_05} S. Stufler, P. Ester, A. Zrenner, and M. Bichler, Phys. Rev. B {\bf 72}, 121301(R) (2005).


\bibitem{Lodahl_05} P. Lodahl, A.P. Mosk, A. Lagendijk, Phys. Rev. Lett. {\bf 95}, 173901 (2005).

\bibitem{Mano_04}T. Mano, R. N\"{o}tzel, G. J. Hamhuis, T. J. Eijkemans, and J. H. Wolter, J. Appl. Phys. {\bf 95}, 109 (2004). 

\bibitem{Lippen_04} T. v. Lippen, R. N\"{o}tzel, G. J. Hamhuis, and J. H. Wolter, Appl. Phys. Lett. {\bf 85}, 118 (2004).

\bibitem{Agranovich_b84} V.M. Agranovich and V.L. Ginzburg,
\textit{Crystal Optics with Spatial Dispersion and Excitons}, 2nd
ed (Springer, New York, 1984).

\bibitem{Zhang_94} W. Zhang and D. F. Walls, \pra \textbf{49}, 3799 (1994).

\bibitem{Yu_b01} P. Y. Yu and M. Cardona,
\textit{Fundumentals of Semiconductors: Physics and Material Properties}, (Springer, Berlin, 1999).




\end{thebibliography}
\end{document}